\documentclass[%
 aip,
 amsmath,amssymb,
 reprint,%
]{revtex4-1}

\usepackage{graphicx}
\usepackage{dcolumn}
\usepackage{bm}

\usepackage[utf8]{inputenc}
\usepackage[T1]{fontenc}
\usepackage{mathptmx}
\usepackage{etoolbox}

\makeatletter
\def\@email#1#2{%
 \endgroup
 \patchcmd{\titleblock@produce}
  {\frontmatter@RRAPformat}
  {\frontmatter@RRAPformat{\produce@RRAP{*#1\href{mailto:#2}{#2}}}\frontmatter@RRAPformat}
  {}{}
}%
\makeatother
\begin{document}

\preprint{AIP/123-QED}

\title[Sample title]{New analytic formulae for memory and prediction functions in reservoir computers with time delays}
\author{Peyton Mullarkey}
 \author{Sarah Marzen}%
 \email{smarzen@natsci.claremont.edu}
\affiliation{ 
Department of Natural Sciences, Pitzer and Scripps College \\
Claremont, CA 91711
}%

\date{\today}

\begin{abstract}
Time delays increase the effective dimensionality of reservoirs, thus suggesting that time delays in reservoirs can enhance their performance, particularly their memory and prediction abilities. We find new closed-form expressions for memory and prediction functions of linear time-delayed reservoirs in terms of the power spectrum of the input and the reservoir transfer function. We confirm this relationship numerically for some time-delayed reservoirs using simulations, including when the reservoir can be linearized but is actually nonlinear. Finally, we use these closed-form formulae to address the utility of multiple time delays in linear reservoirs in order to perform memory and prediction, finding similar results to previous work on nonlinear reservoirs. We hope these closed-form formulae can be used to understand memory and predictive capabilities in time-delayed reservoirs.
\end{abstract}

\maketitle

\begin{quotation}
Reservoir computing promises to aid memory and prediction of input time series, with universal approximation guarantees\cite{grigoryeva2018echo,gonon2019reservoir,gonon2021fading,gonon2019reservoir} and major empirical successes bolstering its claims that it can do so\cite{pathak2018model,walleshauser2022predicting,gauthier2021next}. If properly harnessed, reservoir computers could predict stock prices, the weather, radar signals, naturalistic video, sports game, or really anything\cite{jaeger2001short,maass2002real}. Much work has been done on what reservoir to use and why\cite{ganguli2008memory,schrauwen2008computational,white2004short,marzen2024complexity,marzen2017difference,marzen2021choosing,han2021revisiting,hsu2023strange,hsu2020time,cuchiero2021discrete,hart2024attractor}, including general theorems delineating under what conditions reservoir computers are universal\cite{gonon2021fading,cuchiero2021discrete,gonon2019reservoir}. Still, designing a good efficient reservoir for an application can be difficult and can require a great deal of expert knowledge\cite{lukovsevivcius2012reservoir,yan2024emerging,platt2022systematic}. A major goal of reservoir computing is to design reservoirs so that reservoir computing can compete with Long Short-Term Memory Units\cite{hochreiter1997long} or transformers\cite{vaswani2017attention} in terms of accuracy even when only reduced state space information is available\cite{vlachas2019forecasting}, as reservoir computers are more efficient to train\cite{sheaforecasting,vlachas2019forecasting}. We therefore hope to help understand the computation performed by reservoirs well enough so that design of reservoirs for a particular application-- including, as we address, what time delays to include\cite{duan2023embedding,tavakoli2024boosting,appeltant2011information}-- is straightforward and elegant rather than something you brute-force or learn after years of painstaking research.
\end{quotation}

\section{Introduction}
\label{sec:introduction}

Reservoir computers are simple in concept, as they are essentially recurrent neural networks that are easier to train. To make a reservoir, you just need something whose state responds to input. Over time, that state naturally picks up a memory trace of the input, and this memory trace can be used to remember past inputs, predict future inputs, or reconstruct the dynamical system. (Note that although these goals are related, they are not the same; some memory is required to predict, but some memory can be useless for prediction\cite{marzen2017difference,marzen2016predictive}.) This reservoir state is then fed to a readout layer that uses the state to predict as well as possible. Only the feedforward readout layer is trained. This makes it quite easy and efficient to reconstruct and remember past input and/or predict future input compared to alternatives, as backpropagation through time (which is used to train recurrent neural networks) often has vanishing or exploding gradients\cite{pascanu2013difficulty}.

Anything can be a reservoir, including a collection of neurons or a vat of water\cite{yan2024emerging}. Oftentimes, reservoirs have a number of properties that might aid their ability to process the input, including tuned time delays, tuned weight matrices, or carefully constructed nonlinearities. Even though reservoir computers lack the learning rules (such as backpropagation through time) that characterize the more powerful and thus smaller recurrent neural networks like the Long Short-Term Memory Unit\cite{hochreiter1997long}, they are still able to approximate any time series given to them, as long as they're big enough and the readout is powerful enough\cite{grigoryeva2018echo}. A number of results have shown that reservoir computers have the potential\cite{pathak2018model,walleshauser2022predicting,gauthier2021next} to transform our ability to memorize and predict complex time series signals.

Still, with finite resources, some optimization of the reservoir is often desired. But what to optimize? Roughly speaking, the quality of the reservoir for typical time series tasks is given by the memory function, whose sum is the memory capacity\cite{jaeger2001short,dambre2012information}, and the prediction function, whose sum is the predictive capacity\cite{marzen2017difference}. Much work has been spent on the memory function, with less attention to the prediction function, despite its greater practical implications. Roughly speaking, there is a relationship between the two: You need a certain amount of memory for a desired amount of prediction\cite{marzen2016predictive}. However, it is possible to optimize memory while minimizing prediction in pathological cases\cite{marzen2017difference}.


By trying to understand how time delays (naturally present in most biological and physical systems) in reservoirs affect the memory and prediction function, we hope in this work to take a step towards optimizing the information processing capabilities of reservoirs. Time delays in the reservoir itself, rather than in the readout layer\cite{duan2023embedding}, are an especially powerful tool for reservoirs\cite{appeltant2011information}, as they naturally increase their dimensionality to potentially uncountably infinite, if the time delay is irrational. However, these time delays must be tuned properly for adequate performance in certain test cases, with interesting findings on how multiple time delays are generally preferred\cite{tavakoli2024boosting}.

Here, we study in a minimal model the impact of time delays on the information processing performance of reservoir computers subject to memoryful input. In Sec. \ref{sec:ResultsA}, we provide new closed-form expressions for the memory and prediction functions of linear time-delayed reservoirs. In Sec. \ref{sec:example1}, we confirm that they work on a simple numerical example. In Sec. \ref{sec:ResultsC}, we show that these formulae can even be used to approximate the performance of nonlinear time-delayed reservoirs or on input that is badly understood. In Sec. \ref{sec:ResultsD}, we then use the formulae to help derive some qualitative rules for creating reservoirs that process information as well as possible, with one finding being: that you should have multiple time delays rather than just one. We hope that our new analytic insights will lead to improved understanding, which will yield improved reservoir recipes.

\section{Background}

\subsection{Reservoir computers}

Reservoir computing is a machine learning framework that is used to process and predict time series data. It is a type of artificial neural network designed to predict a system's future outcomes based on its past behavior, coming from echo state networks\cite{jaeger2001short} and liquid state machines\cite{maass2002real}. Unlike standard recurrent neural networks, which train both the reservoir and the output layers as for Long-Short Term Memory Units\cite{hochreiter1997long}, for example, reservoir computers train only the output layers, making them computationally efficient\cite{pascanu2013difficulty} aside from the larger reservoir required to achieve comparable performance to some recurrent neural networks. It operates with a fixed dynamic core, known as the ``reservoir'', and trains only the output connections, simplifying the overall training process compared to recurrent neural networks. 

The fixed reservoir in a reservoir computing system is an internal nonlinear or linear dynamical system that processes and stores information about past inputs. Unlike recurrent neural networks, the internal dynamics of the reservoir are fixed after the initial design and are not modified during training. The fixed design avoids issues like the vanishing gradient problem, which can limit learning in traditional neural networks. The vanishing gradient problem is a challenge that occurs during the training of deep neural networks, especially in architectures with many layers such as (effectively) recurrent neural networks. It happens when the gradients of the loss function, calculated with respect to the weights in the earlier layers, become extremely small as they are propagated backward through thee network during backpropagation.  During backpropagation, the gradients are computed by applying the chain rule to calculate how the weights at each layer contribute to the final loss. This involves multiplying gradients layer by layer. If the activation functions used in the network have derivatives less than 1, successive multiplications across many layers will cause the gradients to shrink exponentially. When the gradients head towards 0, the weights in the earlier layers receive minimal updates during training. This will prevent the network from learning effectively. However, in reservoir computing systems this problem is avoided because weights between the input layer and the reservoir are assigned randomly, and training focuses solely on the output layers. 

In reservoir computers, training is limited to the output layer, which maps the reservoir's hidden states to the desired output. This is achieved using a regularized linear least-squares optimization procedure or linear regression, making training computationally efficient. The reservoir must be large enough to capture the input dynamics, as it does not adapt its internal state during training. 

Time delays have been shown to play a beneficial role in reservoir computing by enhancing both the memory and prediction capabilities\cite{duan2023embedding,tavakoli2024boosting}. Introducing a time delay in a reservoir means incorporating feedback mechanisms or structures that delay the propagation of information within the reservoir. The idea is that time delays allow the reservoir to effectively increase its dimensionality, as potentially an uncountable infinity of reservoir values are required for simulation. Increasing such dimensionality is important for predicting complex time series data. However, selecting the optimal configuration of time delays is critical\cite{tavakoli2024boosting}; improper delay setting, such as matching the delay length to the system's clock cycle, can degrade memory and predictive performance. If the delay length matches the clock cycle, input data may overlap destructively, reducing the reservoir's memory capacity and predictive accuracy. Also, while time delays may improve memory retention, excessive delays might lead to information redundancy, negatively impacting prediction accuracy. 

The inherent non-linearity of certain natural systems allows for physical reservoirs. For example, researchers have demonstrated that water can be a reservoir\cite{fernando2003pattern}. Input signals are introduced through electric motors that create ripples on the water surface, which are then analyzed in the output layer for tasks such as pattern recognition. This example shows the flexibility of reservoir computing in leveraging natural systems for computation. Here, we focus on a theoretical reservoir that can be implemented \emph{in silico}.

Reservoir computing has shown promise in tasks like time-series forecasting, speech recognition, and pattern classification. Despite being less accurate than recurrent neural networks in some cases\cite{}, reservoir computers simplicity and adaptability make it a powerful tool\cite{pathak2018model}. Incorporating time delays offers a way to improve performance, if they are optimized\cite{tavakoli2024boosting}. 

\subsection{Memory function and prediction function}

The memory and prediction function, $m(\tau)$, quantifies the ability of a reservoir computer to remember past inputs for a specific time delay $\tau$. Its value determines how effectively the reservoir state $s(t)$ can encode and recall input signals $x(t)$ from its past (memory function) or predict those in its future (prediction function), which is important for time-series prediction tasks that require temporal dependencies. The memory function for a one-dimensional reservoir is defined by:
\begin{eqnarray}
m(\tau) &=& \frac{\left(\langle x(t+\tau) s(t)\rangle-\langle x(t)\rangle\langle s(t)\rangle\right)^2}{\left(\langle s(t)^2\rangle-\langle s(t)\rangle^2\right) \left(\langle x(t)^2\rangle-\langle x(t)\rangle^2\right)}
\label{eq:1}
\end{eqnarray}
where averages are taken with respect to time $t$. Positive $\tau$ here corresponds to prediction, while negative $\tau$ corresponds to memory, as $m(\tau)$ is simply a squared correlation coefficient between $s(t)$ and $x(t+\tau)$. For reservoirs in which $s$ is of higher-dimension, we have
\begin{eqnarray}
m(\tau) &=& \frac{p_{\tau}^{\top}C^{-1}p_{\tau}}{\sigma_{xx}^2}
\end{eqnarray}
for $p_{\tau}$ the covariance between $s(t)$ and $x(t+\tau)$, $C$ the covariance matrix for $s(t)$, and $\sigma_{xx}^2$ the variance of $x(t)$ \cite{marzen2017difference}. (To see why this is true, recall that in linear regression, the squared correlation coefficient is exactly solvable in terms of covariance.)

The memory capacity $MC$\cite{jaeger2001short} is the sum or integral of $m(\tau)$ over all negative $\tau$, while the predictive capacity $PC$\cite{marzen2017difference} is the sum or integral of $m(\tau)$ over all positive $\tau$. In other words,
\begin{equation}
MC = \int_{-\infty}^0 m(\tau) d\tau
\end{equation}
and
\begin{equation}
PC = \int_0^{\infty} m(\tau) d\tau.
\end{equation}
These capture overall how much memory or predictive capability a reservoir has for a specific input. In general, $MC$ needs to be of a certain value for $PC$ to be large, although it is possible to have arbitrarily high $MC$ and negligible $PC$\cite{marzen2017difference}.

\section{Results}
\label{sec:Results}

\subsection{New expression for memory and prediction functions}
\label{sec:ResultsA}

Recall that the memory function for a one-dimensional reservoir is defined by Eq. \ref{eq:1}. Positive $\tau$ here corresponds to prediction, while negative $\tau$ corresponds to memory, as $m(\tau)$ is simply a squared correlation coefficient between $s(t)$ and $x(t+\tau)$. We will rewrite this in terms of the power spectral density of the input $P(\omega)$ and the transfer function of the linear time-delayed reservoir $H(\omega)$.

In general, we represent a linear reservoir with time-delays to have an evolution equation given by
\begin{equation}
\frac{ds}{dt} = \sum_j K_j s(t-T_j) + v x(t)
\end{equation}
which implies after a Fourier transform that
\begin{eqnarray}
i\omega \hat{s}(\omega) &=& \sum_j K_j e^{-i\omega T_j} \hat{s}(\omega) + v \hat{x}(\omega) \\
\hat{s}(\omega) &=& \frac{v}{i\omega - \sum_j K_j e^{-i\omega T_j}} \hat{x}(\omega),
\end{eqnarray}
yielding a linear transfer function of
\begin{equation}
H(\omega) = \frac{\hat{s}(\omega)}{\hat{x}(\omega)}= \frac{v}{i\omega - \sum_j K_j e^{-i\omega T_j}}
\label{eq:transferfunction1}
\end{equation}
so that
\begin{equation}
\hat{s}(\omega) = H(\omega) \hat{x}(\omega).
\end{equation}
Some equation manipulation, showing that the correlation between reservoir and input is \begin{equation}
\langle x(t+\tau) s(t)\rangle-\langle x(t)\rangle\langle s(t)\rangle =    \frac{1}{2\pi}\int_{-\infty}^{\infty} e^{-i\omega \tau} P(\omega) H(\omega) d\omega,
\end{equation}
that the variance of the input is 
\begin{equation}
\langle x(t)^2\rangle - \langle x(t)\rangle^2 = \frac{1}{2\pi}\int_{-\infty}^{\infty} P(\omega) d\omega,
\end{equation}
and the variance of the reservoir state is
\begin{equation}
 \langle s(t)^2\rangle -\langle s(t)\rangle^2 =   \frac{1}{2\pi}\int_{-\infty}^{\infty} P(\omega) |H(\omega)|^2 d\omega
\end{equation}
where $P(\omega)$ is the power spectral density gives
\begin{equation}
m(\tau) = \frac{\left(\int_{-\infty}^{\infty} e^{-i\omega \tau} P(\omega) H(\omega) d\omega\right)^2}{\int_{-\infty}^{\infty} P(\omega) |H(\omega)|^2 d\omega \int_{-\infty}^{\infty} P(\omega) d\omega}.
\label{eq:answer}
\end{equation}
See Appendix \ref{app:1} for details.


In the more general case of a linear transfer function to an $n$-dimensional linear time-delayed reservoir, we have a similar expression\cite{marzen2017difference}, where now $H(\omega)$ is a vector:
\begin{eqnarray}
m(\tau) &=& \frac{p_{\tau}^{\top}C^{-1}p_{\tau}}{\frac{1}{2\pi}\int_{-\infty}^{\infty}P(\omega)d\omega} \label{eq:general_answer}
\end{eqnarray}
where
\begin{equation}
p_{\tau} = \frac{1}{2\pi}\int_{-\infty}^{\infty} e^{-i\omega \tau} P(\omega) H(\omega) d\omega
\end{equation}
and
\begin{equation}
C = \frac{1}{2\pi}\int_{-\infty}^{\infty} P(\omega) H(\omega) H^{\top}(-\omega) d\omega.
\end{equation}
For details, see Appendix \ref{app:1}.

\subsection{Confirming new analytic expression for memory and prediction function in a simple example}
\label{sec:example1}

We test this expression for the memory and prediction function on a simple input example with an analytically solvable power spectrum. To investigate this, we used a dynamical system that describes the evolution of a stimulus, $x(t)$, defined by the following equations: 
\begin{eqnarray}
\frac{dx}{dt} &=& v \\
\frac{dv}{dt} &=& -k x - \gamma v + D\eta(t)
\end{eqnarray}
where $\eta(t)$ represents stochastic Gaussian noise, $k$ and $\gamma$ are constants that model spring restoring forces and damping and $D$ determines the noise intensity. This system stimulates a noisy, damped harmonic oscillator that serves as the input to the reservoir. This has a power spectrum of
\begin{equation}
P(\omega) = \frac{D^2}{(-\omega^2+k)^2+\omega^2\gamma^2}.
\end{equation}
See Appendix \ref{app:2}. This input has some predictable components, some randomness that made it somewhat hard to predict, and more importantly for our purposes, an analytically-known power spectrum.

\begin{figure}
\centering
\includegraphics[width=0.45\textwidth]{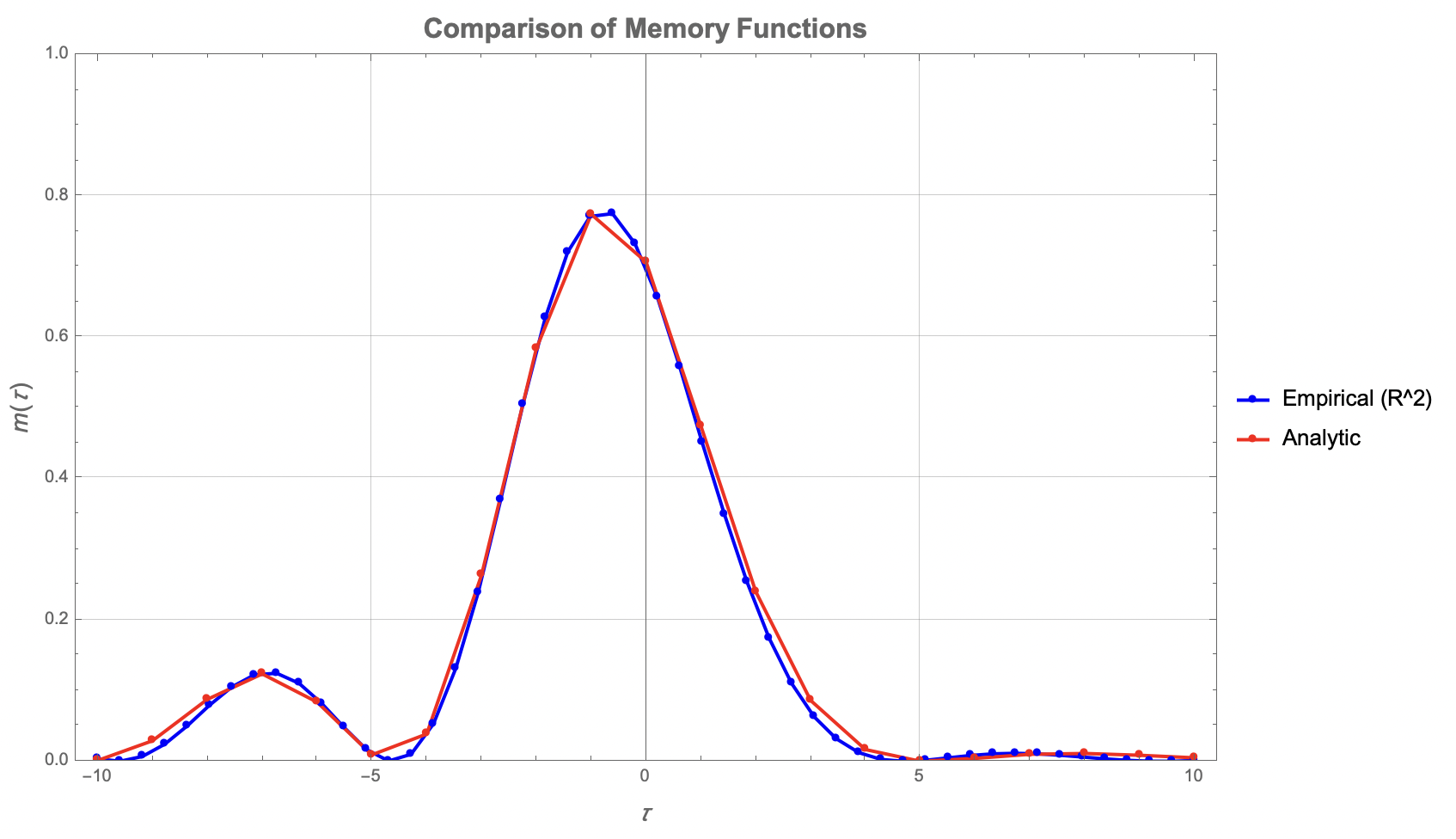}
\caption{The analytic formula for $m(\tau)$ in Eq. \ref{eq:answer}, the memory and prediction function, matches the simulated memory and prediction function from squared correlation coefficients. Reservoir and input has parameters $k = 0.2$, $K = 0.2$, $D = 1$, $\gamma = 0.5$, and with time delays of $1,~2$, and $3$.}
\label{fig:2}
\end{figure}

We use a reservoir with an evolution equation of
\begin{eqnarray}
\frac{ds}{dt} &=& -K \sum_{i=1}^{M} \left( s(t - T_i) - x(t) \right)
\end{eqnarray}
as this reservoir has a fixed point when prediction is achieved, $s(t)=x(t+T_i)$, so a reservoir of this type might enable strong prediction. The transfer function is found via Fourier transform,
\begin{eqnarray}
i \omega \hat{s}(\omega) &=& -K \left( \sum_{j=1}^{M} \hat{s}(\omega) e^{-i \omega T_j} - M \hat{x}(\omega) \right) \\
\hat{s}(\omega) &=& \frac{K M \hat{x}(\omega)}{i \omega + K \sum_{j=1}^{M} e^{-i \omega T_j}}
\end{eqnarray} 
yielding a linear transfer function of
\begin{equation}
H(\omega) = \frac{\hat{s}(\omega)}{\hat{x}(\omega)}= \frac{K M}{i \omega + K \sum_{j=1}^{M} e^{-i \omega T_i}}.
\label{eq:transferfunction2}
\end{equation}
Using Eq. \ref{eq:answer}, Mathematica's NIntegrate produces the same result as a simulation of length $100$ with timestep $dt=0.01$ using the Euler-Marayama equations. See Fig. \ref{fig:2}.

\begin{figure}
\centering
\includegraphics[width=0.45\textwidth]{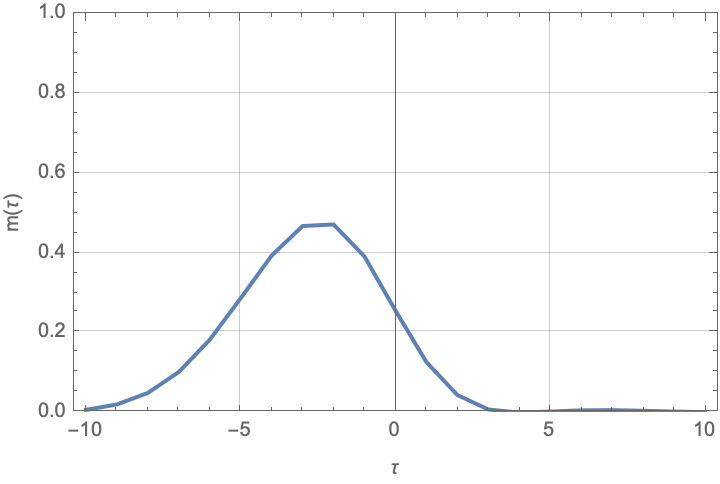}
\caption{The analytic formula for $m(\tau)$ in Eq. \ref{eq:answer}, the memory and prediction function, can handle irrational time delays that would be hard to simulate. Irrational time delays technically lead to an infinite-order reservoir and involve propagating a function rather than a finite number of values. Reservoir and input have parameters $k = 0.2$, $K = 0.2$, $D = 1$, $\gamma = 0.5$, and a time delay of $\pi$.}
\label{fig:3}
\end{figure}

A benefit to using Mathematica's NIntegrate is that one can deal with irrational time delays, which lead to infinite-dimensional reservoirs, up to machine precision, which is difficult using simulation. To simulate reservoirs with irrational time delays requires that you initialize a function on a bounded interval rather than initializing a finite number of values, and then the function must be propagated. This did not appear to change the memory and prediction function much, but it is an advantage to the analytic approach, when it can be employed. See Fig. \ref{fig:3}.

\subsection{Computational benefits from analytic formulae even when reservoir is nonlinear or input is not fully understood}
\label{sec:ResultsC}

Oftentimes, the reservoir is nonlinear, or the input's power spectrum is not entirely well-known. We show how these closed-form expressions can still be used to approximately find the memory and prediction function's behavior.

\subsubsection{Nonlinear reservoirs}

\begin{figure}
\centering
\includegraphics[width=0.45\textwidth]{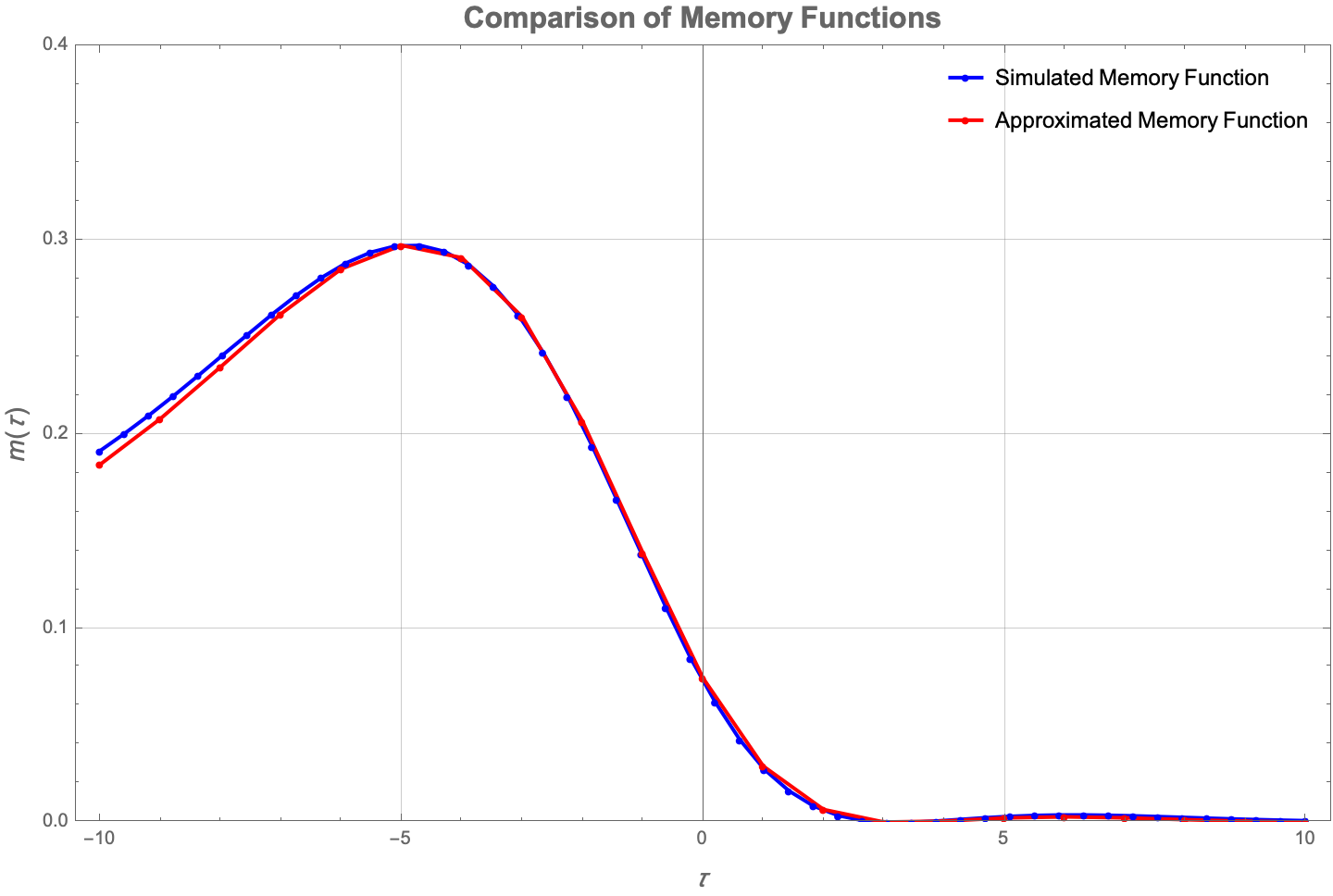}
\caption{Memory and prediction function calculated as though the reservoir is approximately linear using these closed-form expressions in Eq. \ref{eq:1C} combined with Eq. \ref{eq:answer} (``Approximated Memory Function'') and via simulations directly from squared correlation coefficients (``Simulated Memory Function''). Reservoir and input use parameters of $k = 0.2$, $K = 0.2$, $D = 1$, $\gamma = 0.5$, $\beta=0.05$ and time delays of $1$, $2$, and $3$. Note that unlike in previous figures, the memory function is maximal at $\tau\approx -5.0$ rather than $\tau\approx -2.5$, showing that the location of the peak in the memory function is a property of the reservoir in addition to a property of the input.}
\label{fig:4}
\end{figure}

Linearizing the system, if the system has some form of fixed point behavior, may \emph{still} yield an understanding of memory and prediction using these closed-form expressions. For instance, if
\begin{eqnarray}
\frac{ds}{dt} &=& -K \sum_{i=1}^{M} \tanh\left( \beta(s(t - T_i) - x(t) )\right)
\label{eq:1A}
\end{eqnarray}
by using $\beta\ll 1$ and pretending that the reservoir is the \emph{linearized} version,
\begin{eqnarray}
\frac{ds}{dt} &=& -K\beta \sum_{i=1}^{M} (s(t - T_i) - x(t) ),
\label{eq:1B}
\end{eqnarray}
so that its transfer function is approximately
\begin{equation}
H(\omega) \approx \frac{K \beta M}{i \omega + K\beta \sum_{j=1}^{M} e^{-i \omega T_i}}
\label{eq:1C}
\end{equation}
We compare simulated (so more exact) memory and prediction functions in Fig. \ref{fig:4} using Eq. \ref{eq:1A} to the approximated analytic formulae using Mathematica NIntegrate with equation \ref{eq:1C} for $\beta=0.05$ using the aforementioned mass-on-a-spring stimulus so that the input power spectrum was known perfectly. We find good agreement, despite the nonlinearity. As expected, the analytic formulae are useful as long as linearization is appropriate-- which may likely only hold when the autonomous version of the reservoir has a fixed point rather than oscillations or chaos.

\subsubsection{Approximated power spectrum of the input}

Sometimes the power spectrum of an input is known only approximately, leading to errors in using the estimates of the memory and prediction function using Eq. \ref{eq:answer}. In particular, an error $\delta P(\omega)$ in the power spectrum leads to an error for the one-dimensional reservoir's memory and prediction function:
\begin{eqnarray}
\frac{\delta m(\tau)}{m(\tau)} &=& 2\frac{\int_{-\infty}^{\infty} e^{-i\omega \tau} \delta P(\omega) H(\omega) d\omega}{\int_{-\infty}^{\infty}e^{-i\omega \tau} P(\omega) H(\omega) d\omega} -\frac{\int_{-\infty}^{\infty} \delta P(\omega) |H(\omega)|^2 d\omega}{\int_{-\infty}^{\infty} P(\omega) |H(\omega)|^2 d\omega} \nonumber \\
&& - \frac{\int_{-\infty}^{\infty} \delta P(\omega)}{\int_{-\infty}^{\infty} P(\omega) d\omega}
\label{eq:corrections}
\end{eqnarray}
to first order in $\delta P(\omega)$. See Appendix \ref{app:4}.
Thus, even if your power spectrum is not quite right, the analytic expression provided in Eq. \ref{eq:answer} can still provide a rough guide as to the features of the memory and prediction function, with corrections given by Eq. \ref{eq:corrections} that decrease as $\delta P(\omega)$ decreases.

Unfortunately, numerical estimation of the power spectrum combined with approximate evaluation of the integral involved in $m(\tau)$ is fraught with error. One would hope that Scipy's built-in packages or other numerical packages could estimate the input's power spectrum from the input time series, and that this numerically-obtained power spectrum could then be fed into the equations for the memory and prediction functions, where the oscillatory integrals could be numerically approximated. However, when we tried to estimate the power spectrum of the noisy simple harmonic oscillator using Scipy's Signal package, it was significantly different than the ground truth analytic answer. Then, direct Riemann summation to estimate the numerical integrals for the memory and prediction functions resulted in negative memory function and prediction function values.

\subsection{Investigating utility of multiple time delays in memory and prediction for an example reservoir}
\label{sec:ResultsD}

One of the key benefits of closed-form expressions is the insight you are able to attain. We now turn our interest to reservoirs with multiple time delays, asking if there is an advantage to multiple time delays even in simple linear reservoirs\cite{tavakoli2024boosting}.

\begin{figure}
\centering
\includegraphics[width=0.45\textwidth]{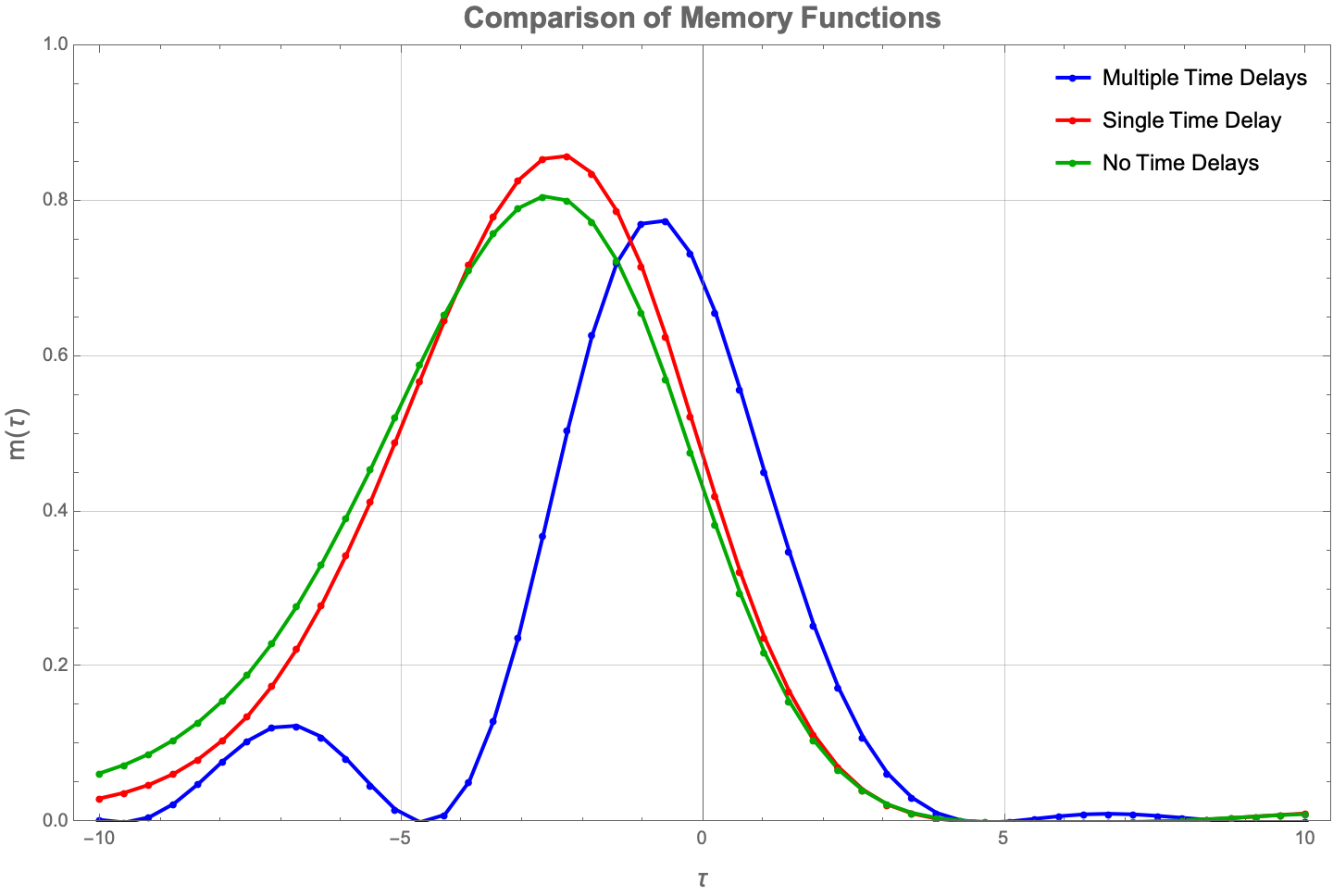}
\includegraphics[width=0.45\textwidth]{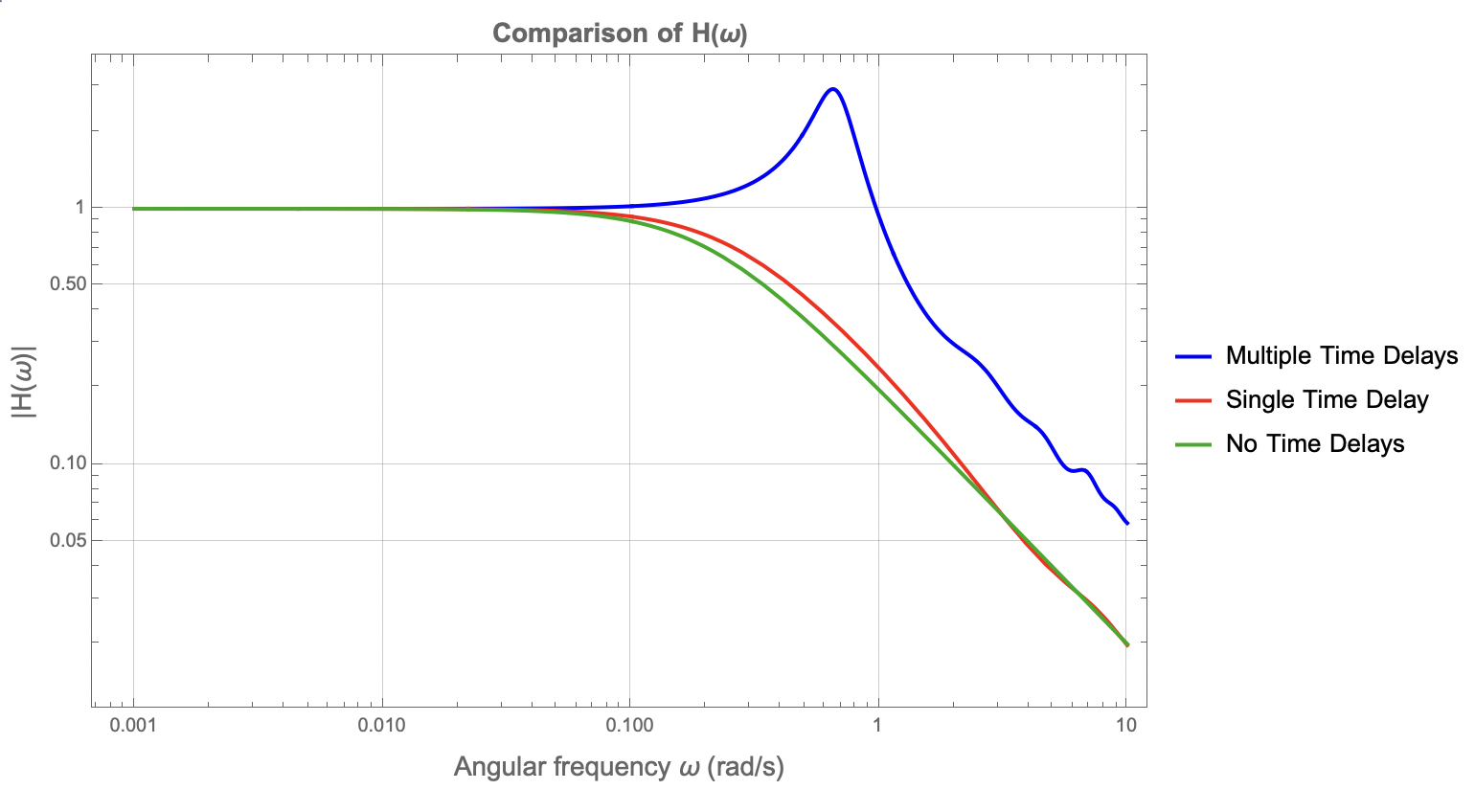}
\caption{(Top) The memory and prediction function of reservoirs with varying numbers of time delays for the same input and same ``stiffness'' $k$. (Bottom) Their transfer functions, shown with a linear scale on the $y$-axis to accentuate the differences between the reservoirs. For both, with parameters $k = 0.2$, $K = 0.2$, $D = 1$, $\gamma = 0.5$, we evaluate the following reservoirs. For multiple time delays, we have time delays of $1$, $2$, and $3$; for a single time delay, $T=1$; for no time delays, $T=0$.}
\label{fig:5}
\end{figure}

It is thought that multiple time delays can provide huge memory advantages and predictive advantages\cite{tavakoli2024boosting}. To test this in our simple setup, we use a reservoir with an evolution equation of
\begin{eqnarray}
\frac{ds}{dt} &=& -K \sum_{i=1}^{M} \left( s(t - T_i) - x(t) \right)
\end{eqnarray}
and the input of a mass on a spring:
\begin{eqnarray}
\frac{dx}{dt} &=& v \\
\frac{dv}{dt} &=& -k x - \gamma v + D\eta(t).
\end{eqnarray}
Based on our investigations, aided by the analytic formulae, we can understand whether or not multiple time delays lead to increased memory and prediction, as obtained previously for a different input and reservoir \cite{tavakoli2024boosting}. From visual inspection of Fig. \ref{fig:5}(top), multiple time delays lead to slight increases in memory and prediction that could become more pronounced by using even more time delays. One time delay still provides an advantage in memory and prediction over no time delays.

For this input and these reservoirs, time delays do not need to be tuned precisely, so gains and decreases in memory and prediction as time delays are added and varied is not as drastic as they were in other reservoirs \cite{tavakoli2024boosting}. This is because the reservoir is not naturally oscillatory such that the wrong time delays destroy the ability to learn new information about the input\cite{tavakoli2024boosting}.

The difference in memory and prediction for each of these reservoirs can be understood via the difference in transfer functions. All of these transfer functions constitute low-pass filters, but by adding multiple time delays, the low-pass filters acquire oscillatory components. See Fig. \ref{fig:5}(bottom). It appears that the more oscillations present in the transfer function, the higher the memory, according to Fig. \ref{fig:5}(top).

\section{Discussion}

What we have described is an analysis of linear filters that are disguised as reservoirs. Time delays in the reservoirs turn into oscillatory terms in the denominator of the transfer function of the filter in Eq. \ref{eq:transferfunction1} and Eq. \ref{eq:transferfunction2}, leading to sometimes-observable oscillations in the magnitude of the transfer function itself as in Fig. \ref{fig:5}(bottom). (Note that oscillations in the transfer function's denominator are not the same as oscillatory behavior in the reservoir.) For such filters, the sometimes hard-to-estimate memory and prediction functions can be written in terms of the transfer function of the reservoir itself and the power spectrum of the input. In a simple case, these closed-form expressions match simulations and provide an alternate route towards estimating the memory and predictive capabilities of reservoirs. They even match simulations somewhat when the reservoirs are nonlinear, but can be linearized. When power spectrum of the input must be estimated from data, using these closed-form expressions is quite difficult, due to difficulties in accurate estimation of the power spectrum and also difficulties in numerical integration. Using these formulae show that indeed, multiple time delays do provide an advantage in memory and prediction for reservoirs even in the linear case, echoing previous work in specific nonlinear cases\cite{tavakoli2024boosting}.

This may at first seem to provide a new route to optimize reservoirs, and in a way it does-- but the connection is subtle. As transfer functions describe only linear reservoirs exactly, these closed-form expressions can be used to compute in a new way the memory and prediction functions of reservoirs; but the predictive performance of linear reservoirs is upper-bounded by the Wiener filter. As such, we do not aim to optimize the transfer function of \emph{linear} reservoirs, and instead we envision a different use for these formulae. We hope that the closed-form expression can be analyzed beyond what we have done in this paper so that we gain insight into what kinds of weight matrices and what kinds of time delays (and how many of them) contribute to maximal memory and/or prediction. Whether or not memory or prediction is desired is entirely up to the user, but the work presented here allows for maximization of both.

In other words, following in the tradition of previous theoretical work\cite{white2004short,ganguli2008memory,marzen2017difference,hsu2020time,hsu2023strange,marzen2021choosing,dambre2012information}, we hope that analysis of simple reservoirs will provide new insight into how to optimize more complicated reservoirs beyond what currently exists\cite{yan2024emerging}. This linearization approach has worked in the past for very complicated nonlinear systems, leading even to a new initialization approach for the training of artificial neural networks\cite{saxe2013exact}.

Another way of viewing this analytic contribution is that we have shown that linear time-delayed reservoirs are all low-pass filters of varying types. The question that remains is simply: what kinds of low-pass filters do you want for a given input power spectrum? Can you make a reservoir that is a bandpass or high-pass filter? Time delays in frequency space provide a clear qualitative change in filtering properties by allowing for oscillations in the transfer function. Can these oscillations be optimized for classes of input, in addition to optimizing the weight matrix recipes that most workers spend time on\cite{lukovsevivcius2012reservoir,yan2024emerging,platt2022systematic}?

Looking to the future, we can hope that even highly nonlinear time-delayed reservoirs can be, to first order, thought of as having a transfer function that explains its filtering properties, in the same way that receptive fields in neuroscience can explain some fraction of what neurons actually do despite being simplistic models of neural activity\cite{olshausen2006other}. And as a result, we hope that any insights gleaned into ideal weight matrix constructions and time delays by using the formulae here can be used to determine new reservoir recipes for even nonlinear reservoirs\cite{platt2022systematic}.

\begin{acknowledgments}
We wish to acknowledge helpful comments from John Milton. We would also very much like to thank Daniel Lichtblau for all the help with catching errors in previous version's code and analytics.
\end{acknowledgments}

\section*{Data Availability Statement}

The data that support the findings of this study are openly available in Google Colab using Python \cite{Python} at https://shorturl.at/jykmQ. 

\appendix

\begin{widetext}

\section{Relating memory and prediction function to the power spectrum and transfer function}
\label{app:1}

We start by showing that
\begin{eqnarray}
\langle x(t)^2\rangle-\langle x(t)\rangle^2 &=& \frac{1}{2\pi}\int_{-\infty}^{\infty} P(\omega) d\omega \\
\langle s(t)^2\rangle-\langle s(t)\rangle^2 &=& \frac{1}{2\pi} \int_{-\infty}^{\infty} P(\omega) |H(\omega)|^2 d\omega \\
\langle x(t+\tau) s(t)\rangle-\langle x(t)\rangle\langle s(t)\rangle &=& \frac{1}{2\pi} \int_{-\infty}^{\infty} e^{-i\omega \tau} P(\omega) H(\omega) d\omega
\end{eqnarray}
Without loss of generality, we take the input signal and zero-mean it, so that $\langle x\rangle=0$.
To tackle the first one, leading to an expression for the variance, we note that $x(t) = \frac{1}{2\pi}\int_{-\infty}^{\infty}\hat{x}(\omega) e^{i\omega t} d\omega$
\begin{eqnarray}
\langle x(t)^2\rangle-\langle x(t)\rangle^2 &=& \lim_{T\rightarrow\infty} \frac{1}{2T} \int_{-T}^{T} x(t)^2 dt \\
&=& \lim_{T\rightarrow\infty} \frac{1}{2T} \int_{-T}^{T} \left(\frac{1}{2\pi}\int_{-\infty}^{\infty}\hat{x}(\omega) e^{i\omega t} d\omega\right) \left(\frac{1}{2\pi}\int_{-\infty}^{\infty}\hat{x}(\omega') e^{i\omega' t} d\omega'\right) dt \\
&=& \lim_{T\rightarrow\infty} \frac{1}{2T} \frac{1}{4\pi^2} \int_{-\infty}^{\infty} d\omega \int_{-\infty}^{\infty} d\omega' \int_{-T}^{T} e^{i(\omega+\omega')t} \hat{x}(\omega)\hat{x}(\omega') dt.
\end{eqnarray}
We take the limit as $T\rightarrow\infty$:
\begin{eqnarray}
\langle x(t)^2\rangle &\rightarrow& \frac{1}{2T}\frac{1}{4\pi^2} \int_{-\infty}^{\infty} \int_{-\infty}^{\infty} \left(\int_{-\infty}^{\infty} e^{i(\omega+\omega')t} dt\right) \hat{x}(\omega)\hat{x}(\omega') d\omega d\omega' \\
&=& \frac{1}{2\pi}\frac{1}{2T}\int_{-\infty}^{\infty} \int_{-\infty}^{\infty} \delta(\omega+\omega') \hat{x}(\omega) \hat{x}(\omega') d\omega d\omega' \\
&=& \frac{1}{2\pi} \int_{-\infty}^{\infty} P(\omega) d\omega.
\end{eqnarray}
We then tackle
\begin{eqnarray}
\langle x(t+\tau) s(t)\rangle &\rightarrow& \frac{1}{2T}\int_{-\infty}^{\infty} x(t+\tau) s(t) dt \\
&=& \frac{1}{4\pi^2} \frac{1}{2T}\int_{-\infty}^{\infty} \left(\int_{-\infty}^{\infty} \hat{x}(\omega) e^{i\omega (t+\tau)} dt\right) \left(\int_{-\infty}^{\infty} \hat{s}(\omega') e^{i\omega' t}d\omega'\right) dt \\
&=& \frac{1}{4\pi^2} \frac{1}{2T}\int_{-\infty}^{\infty} \int_{-\infty}^{\infty} \hat{x}(\omega) \hat{s}(\omega') e^{i\omega\tau} \int_{-\infty}^{\infty} e^{i(\omega+\omega')t} dt d\omega d\omega' \\
&=& \frac{1}{2\pi} \frac{1}{2T}\int_{-\infty}^{\infty} \int_{-\infty}^{\infty} \hat{x}(\omega) H(\omega') \hat{x}(\omega') e^{i\omega \tau} \delta(\omega+\omega') d\omega d\omega' \\
&=& \frac{1}{2\pi} \int_{-\infty}^{\infty} P(\omega) H(-\omega) e^{i\omega \tau} d\omega \\
&=& \frac{1}{2\pi} \int_{-\infty}^{\infty} P(\omega) H(\omega) e^{-i\omega \tau} d\omega.
\end{eqnarray}
And finally, for a linear system with no bias, $\langle s\rangle = 0$, and so we merely have by the same manipulations as we did for $\langle x(t)^2\rangle$:
\begin{eqnarray}
\left(\langle s(t)^2\rangle-\langle s(t)\rangle^2\right) &\rightarrow& \frac{1}{2\pi} \frac{1}{2T}\int_{-\infty}^{\infty} \hat{s}(\omega) \hat{s}(-\omega) d\omega \\
\left(\langle s(t)^2\rangle-\langle s(t)\rangle^2\right)&\rightarrow& \frac{1}{2\pi}\frac{1}{2T} \int_{-\infty}^{\infty} H(\omega) \hat{x}(\omega) H(-\omega) \hat{x}(-\omega) d\omega \\
&=& \frac{1}{2\pi} \int_{-\infty}^{\infty} |H(\omega)|^2 P(\omega) d\omega.
\end{eqnarray}
In the more general case, these formulae are straightforward to extend. If $s$ is actually a vector, such that the transfer function is a vector, the entire derivation for $\langle x(t+\tau) s(t)\rangle$ and $\langle s(t)s(t)^{\top}\rangle$ carry over straightforwardly.

\section{Power spectrum of mass on a spring}
\label{app:2}

From the evolution equations,
\begin{eqnarray}
\frac{d^2x}{dt^2} &=& -kx - \gamma \frac{dx}{dt} + D\eta(t)
\end{eqnarray}
we find
\begin{eqnarray}
-\omega^2 \hat{x} &=& - k\hat{x} - i\omega\gamma\hat{x}+D\mathcal{F}[\eta] \\
\hat{x} &=& \frac{D\mathcal{F}[\eta]}{-\omega^2 + k+i\gamma\omega} \\
P(\omega) &=& \frac{D^2}{(k-\omega^2)^2+\gamma^2\omega^2}
\end{eqnarray}
as desired.

\section{Error analysis of memory function}
\label{app:4}

The numerator of $m(\tau)$ is altered to
\begin{eqnarray}
\left(\int_{-\infty}^{\infty} e^{-i\omega\tau} H(\omega) (P+\delta P)(\omega) d\omega\right)^2 &=& \left(\int_{-\infty}^{\infty} e^{-i\omega\tau} H(\omega) P(\omega) d\omega\right)^2 + 2 \left(e^{-i\omega\tau} H(\omega) P(\omega) d\omega\right) \left(e^{-i\omega\tau} H(\omega) \delta P(\omega) d\omega\right) \nonumber \\
&& + O(\delta P^2)
\end{eqnarray}
while the denominator terms are altered to
\begin{eqnarray}
\int_{-\infty}^{\infty} (P+\delta P)(\omega) d\omega &=& \int_{-\infty}^{\infty} P(\omega) d\omega + \int_{-\infty}^{\infty} \delta P (\omega) d\omega
\end{eqnarray}
and
\begin{eqnarray}
\int_{-\infty}^{\infty} |H(\omega)|^2 (P+\delta P)(\omega) d\omega &=& \int_{-\infty}^{\infty} |H(\omega)|^2 P(\omega) d\omega + \int_{-\infty}^{\infty} |H(\omega)|^2 \delta P (\omega) d\omega
\end{eqnarray}
which implies
\begin{eqnarray}
(m+\delta m)(\tau) &=& \frac{\left(\int_{-\infty}^{\infty} e^{-i\omega\tau} H(\omega) P(\omega) d\omega\right)^2 + 2 \left(\int_{-\infty}^{\infty} e^{-i\omega\tau} H(\omega) P(\omega) d\omega\right) \left(\int_{-\infty}^{\infty} e^{-i\omega\tau} H(\omega) \delta P(\omega) d\omega\right)}{\left(\int_{-\infty}^{\infty} P(\omega) d\omega + \int_{-\infty}^{\infty} \delta P (\omega) d\omega\right)\left(\int_{-\infty}^{\infty} |H(\omega)|^2 P(\omega) d\omega + \int_{-\infty}^{\infty} |H(\omega)|^2 \delta P (\omega) d\omega\right)} + O(\delta P^2) \\
&=& \frac{\left(\int_{-\infty}^{\infty} e^{-i\omega\tau} H(\omega) P(\omega) d\omega\right)^2}{\int_{-\infty}^{\infty} P(\omega) d\omega\int_{-\infty}^{\infty} |H(\omega)|^2 P(\omega)} \Big(1+\frac{2 \left(\int_{-\infty}^{\infty}e^{-i\omega\tau} H(\omega) P(\omega) d\omega\right) \left(\int_{-\infty}^{\infty}e^{-i\omega\tau} H(\omega) \delta P(\omega) d\omega\right)}{\left(\int_{-\infty}^{\infty} e^{-i\omega\tau} H(\omega) P(\omega) d\omega\right)^2}-\frac{\int_{-\infty}^{\infty} \delta P(\omega) d\omega}{\int_{-\infty}^{\infty} P(\omega) d\omega} \nonumber \\
&& - \frac{\int_{-\infty}^{\infty} \delta P(\omega) |H(\omega)|^2 d\omega}{\int_{-\infty}^{\infty} P(\omega) |H(\omega)|^2 d\omega}\Big)  + O(\delta P^2) \\
\frac{(m+\delta m)(\tau)}{m(\tau)}&=& 1+\frac{2 \left(\int_{-\infty}^{\infty}e^{-i\omega\tau} H(\omega) P(\omega) d\omega\right) \left(\int_{-\infty}^{\infty}e^{-i\omega\tau} H(\omega) \delta P(\omega) d\omega\right)}{\left(\int_{-\infty}^{\infty} e^{-i\omega\tau} H(\omega) P(\omega) d\omega\right)^2}-\frac{\int_{-\infty}^{\infty} \delta P(\omega) d\omega}{\int_{-\infty}^{\infty} P(\omega) d\omega} - \frac{\int_{-\infty}^{\infty} \delta P(\omega) |H(\omega)|^2 d\omega}{\int_{-\infty}^{\infty} P(\omega) |H(\omega)|^2 d\omega} \nonumber \\
&&+ O(\delta P^2) \\
\frac{\delta m(\tau)}{m(\tau)} &=& \frac{2\int_{-\infty}^{\infty} e^{-i\omega\tau} H(\omega) P(\omega) d\omega}{\int_{-\infty}^{\infty} e^{-i\omega\tau} H(\omega) P(\omega) d\omega}-\frac{\int_{-\infty}^{\infty} \delta P(\omega) d\omega}{\int_{-\infty}^{\infty} P(\omega) d\omega} - \frac{\int_{-\infty}^{\infty} \delta P(\omega) |H(\omega)|^2 d\omega}{\int_{-\infty}^{\infty} P(\omega) |H(\omega)|^2 d\omega} \nonumber \\
&&+ O(\delta P^2)
\end{eqnarray}
using $\frac{1}{1+\frac{\delta x}{x}} \approx 1-\frac{\delta x}{x}$ and $(1+\frac{\delta x}{x})(1-\frac{\delta y}{y})(1-\frac{\delta z}{z}) = 1+\frac{\delta x}{x}-\frac{\delta y}{y}-\frac{\delta z}{z}$ to first order. This then implies the formula in the main text.

\end{widetext}

\bibliography{aipsamp}

\end{document}